\documentstyle[editedvolume]{crckapb} 

\begin{opening}
\title{Deep Fields: The Faint sub-mJy and microJy Radio Sky} 
\subtitle{- A VLBI Perspective}

\author{M.A. Garrett} 
\institute{Joint Institute for VLBI in Europe\\
           Postbus 2, 7990 AA Dwingeloo, The Netherlands}
\end{opening}

\runningtitle{The Faint Radio Sky}

\begin{document}

% The \begin{document} command comes after the \end{opening}
% command.

% \abstract{This is the abstract} 

\section{Introduction}

Until recently, VLBI targets have been drawn almost exclusively from
the brightest and most compact radio sources in the sky, with typical
flux densities well in excess of a few tens of mJy. These sources are
predominantly identified with Active Galactic Nuclei (AGN), located at
cosmological distances ($z \sim 1$). Exotic but also rather rare, these
luminous AGN systems have been studied in great detail by VLBI over the
last 3 decades, producing many front-line discoveries along the way
(see this volume and references therein).

However, in recent years, connected arrays (such as the VLA, WSRT, ATCA
and MERLIN) have also began to focus a significant fraction of their
time towards understanding the nature of the faint radio sky ---
sometimes observing the same field for many days or even weeks at a
time.  At these microJy noise levels, the radio sky literally ``lights
up'', and a new population of vigorous star forming galaxies begin to
dominate the radio source counts (Fomalont et al. 1997, Richards et al.
1998, Muxlow et al. 1999, Richards 2000, Garrett et al. 2000a, Norris
et al. 2000). For many astronomers (usually radio astronomers!) it
comes as some surprise that a well calibrated VLBI array, composed of
the largest telescopes in the world, can also contribute to our
understanding of this sub-mJy and microJy radio source population.
Nevertheless, these are the facts, as recently demonstrated by the
simultaneous detection of 3 sub-mJy radio sources in the Hubble Deep
Field North (HDF-N) by the European VLBI Network (Garrett et al. 2001).

In this lecture I will attempt to summarise what is currently known
about the general properties of the faint sub-mJy and microJy radio
source population, as determined from deep multi-wavelength studies of
the HDF-N. In particular, I will try to provide a VLBI perspective,
describing the first deep, wide-field, VLBI pilot observations of
the HDF, together with a summary of the main results. The role VLBI can
play in future high resolution studies of faint radio sources will also
be addressed. 

\begin{figure}
\vspace{7cm}  
\includegraphics{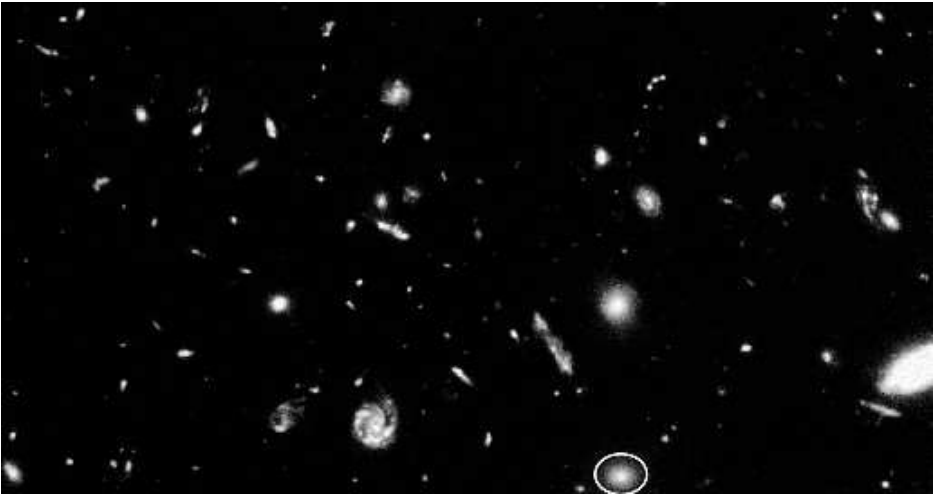}
\caption{Close-up of a small region of the HDF-N (Williams et
  al. 1996). More distant galaxies also tend to be more morphologically
  disturbed (NASA/STScI). One of the radio sources (VLA~J123644+621133)
  detected by the European VLBI Network is circled (see section
  \ref{EVN}).}
\label{hdf_zoom} 
\end{figure}

\section{The Hubble Deep Field and Galaxy Formation} 

The Hubble Deep Field North (HDF-N) is one of the most
important, publicly available resources ever generated in the history of
astronomy (see Ferguson, Dickinson \& Williams 2000 and references
therein). What initially appeared to be a small, undistinguished
$2.5^{\prime} \times 2.5^{\prime}$ patch of the celestial sphere was
transformed during a 10-day long Hubble Space Telescope (HST) deep
field integration (Williams et al. 1996). 

The key advance of the HDF images was not only the depth which they
reached ($I < 29^{m}$) but perhaps more crucially, the superb angular
resolution that accompanied them. On inspection it immediately became
clear (see Fig.~\ref{hdf_zoom}), that the most distant galaxies in the
field were also the most morphologically disturbed.  For example,
familiar ``grand design'' spirals observed locally, all but disappear
beyond $z \sim 0.3$. In terms of morphology, these distant, disturbed
systems are most akin to nearby Ultra Luminous Infrared
Galaxies (ULIG) and interacting starburst systems (Abraham \& van den
Bergh 2001).

\subsection{Multi-wavelength follow-up observations of the HDF} 

\begin{figure}
\vspace{3.5cm}  
\includegraphics{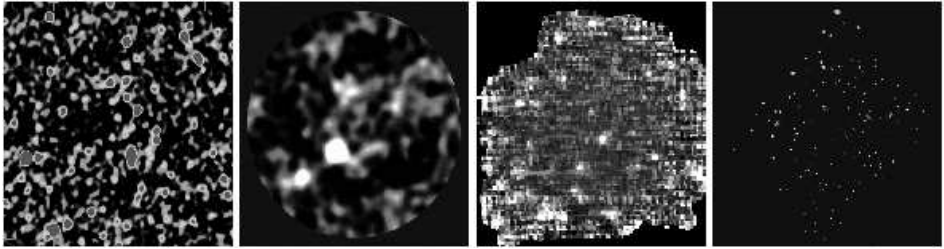}
\caption{Multi-wavelength images of the HDF-N in the radio (Garrett et
  al. 2000a -- includes the HFF), sub-mm (Hughes et al. 1998),
  Mid-IR (Rowan-Robinson et al.  1997 -- includes part of the HFF)
  and x-rays (Brandt et al. 2001 -includes the HFF).}
\label{mband}
\end{figure}

The HDF represented a huge investment in HST observing time, and this
was matched by an equally ambitious multi-wavelength follow-up
campaign. The latter included every part of the e-m spectrum accessible
via both ground and space based instrumentation. In many cases,
extremely long integrations produced the very deepest view of the
Universe in a given wave-band. In addition, extensive efforts were made
to obtain spectroscopic and photometric redshifts of sources in the
field (including the adjacent Hubble Flanking Fields -- HFF). A
pictorial summary of a limited subset of these multi-wavelength follow-up
observations is shown in Fig~\ref{mband}.

It comes as no surprise that the source counts in the HST images (both
in the optical and Near-IR) are considerably greater than that observed
at other wavelengths. What is surprising, however, is that despite this
fact, a considerable fraction (10-20\%) of the faint radio, Mid-IR and
x-ray detections, appear to be heavily obscured in the optical i.e.
they have no obvious optical counterparts (R$ > 25^{m}$).  An even
larger fraction of the faint SCUBA source population fall into this
category, suggesting that the sub-mm observations reveal a completely
different (unobscured) view of the high redshift, dusty, star-forming
Universe (Hughes et al. 1998). We will return to the nature of these
faint sub-mm sources in section \ref{scuba_ref}.

\section{Deep Radio Imaging of the HDF} 
\label{WSRT} 

In order to detect even a handful of radio sources in the HDF-N, noise
levels of a few microJy must be achieved. These in turn require
integration times ranging from a few days -- in the case of the WSRT and
VLA, to many days in the case of MERLIN. 

\begin{figure}
\vspace{8cm}  
\includegraphics{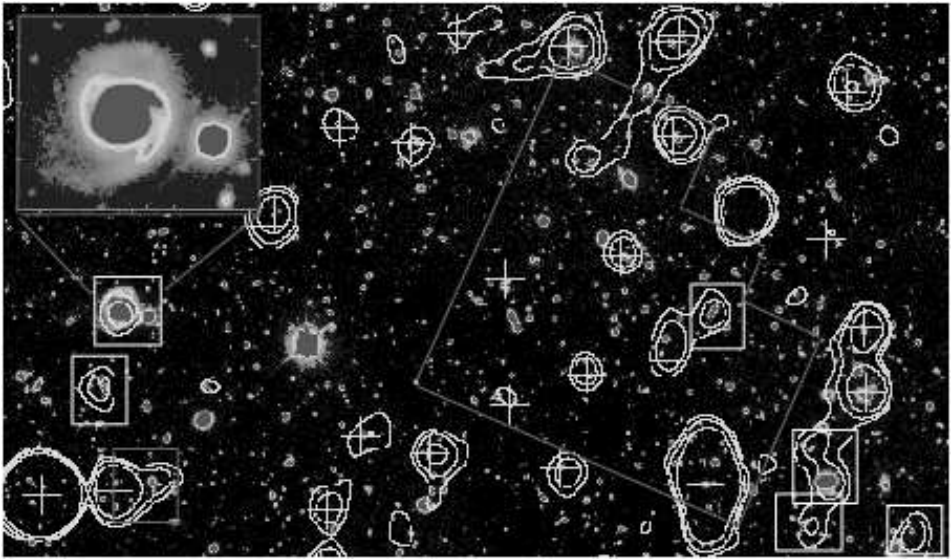}
\caption{The WSRT 1.4~GHz contour map of the HDF (see incomplete
  rotated square) and part of the HFF superimposed upon
  a deep CFHT optical image of the HDF region made by the
  Canada-France-Hawaii telescope (courtesy Amy Barger). Crosses
  indicate previously known VLA detections, boxes indicate new WSRT
  detections. The detection of a nearby, extended star forming galaxy
  is highlighted (upper left).}
\label{wsrt-im} 
\end{figure}

\subsection{VLA 8.3 GHz, VLA-MERLIN \& WSRT 1.4 GHz observations} 

\begin{figure}
\vspace{5cm}  
\includegraphics{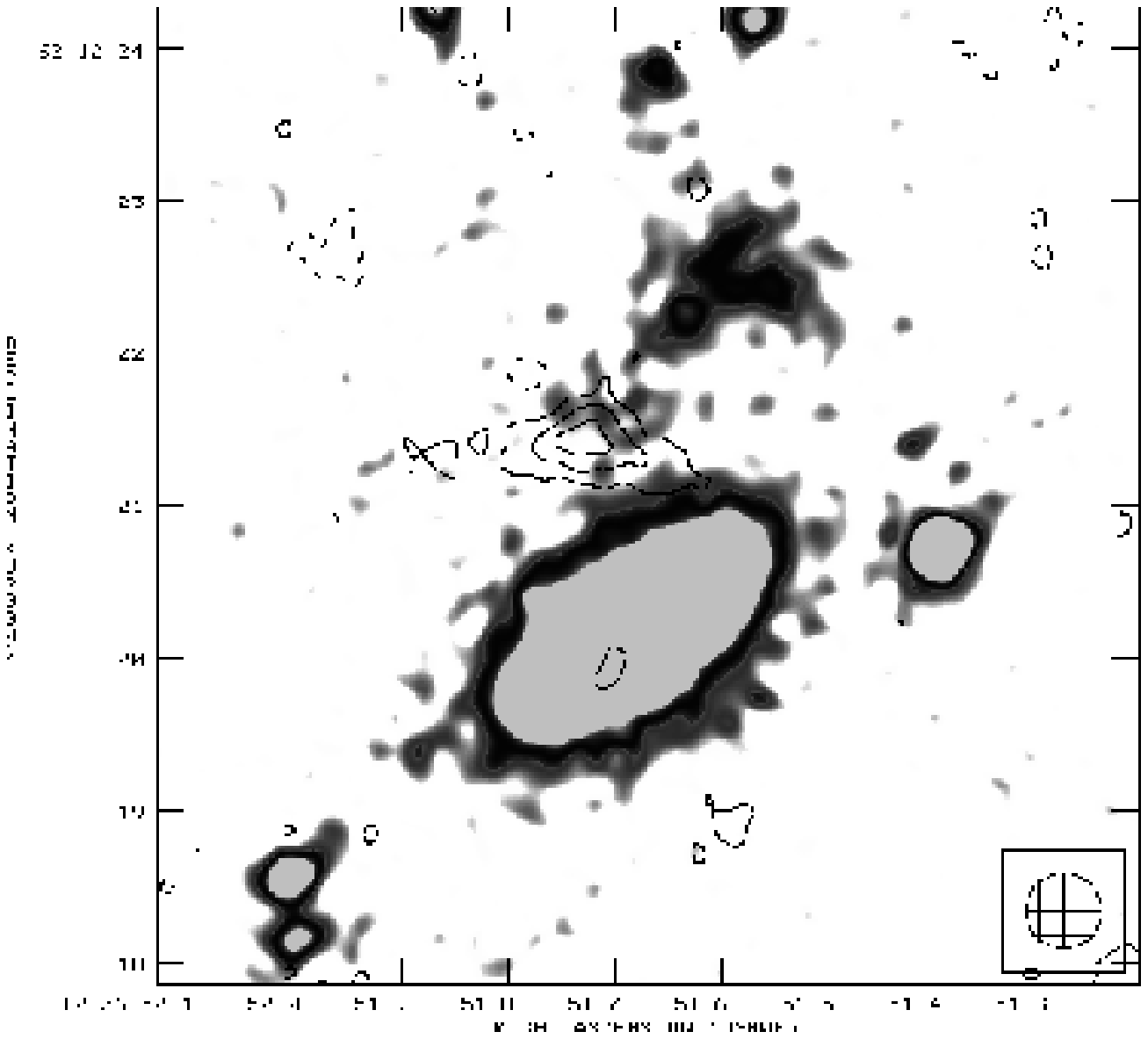}
\includegraphics{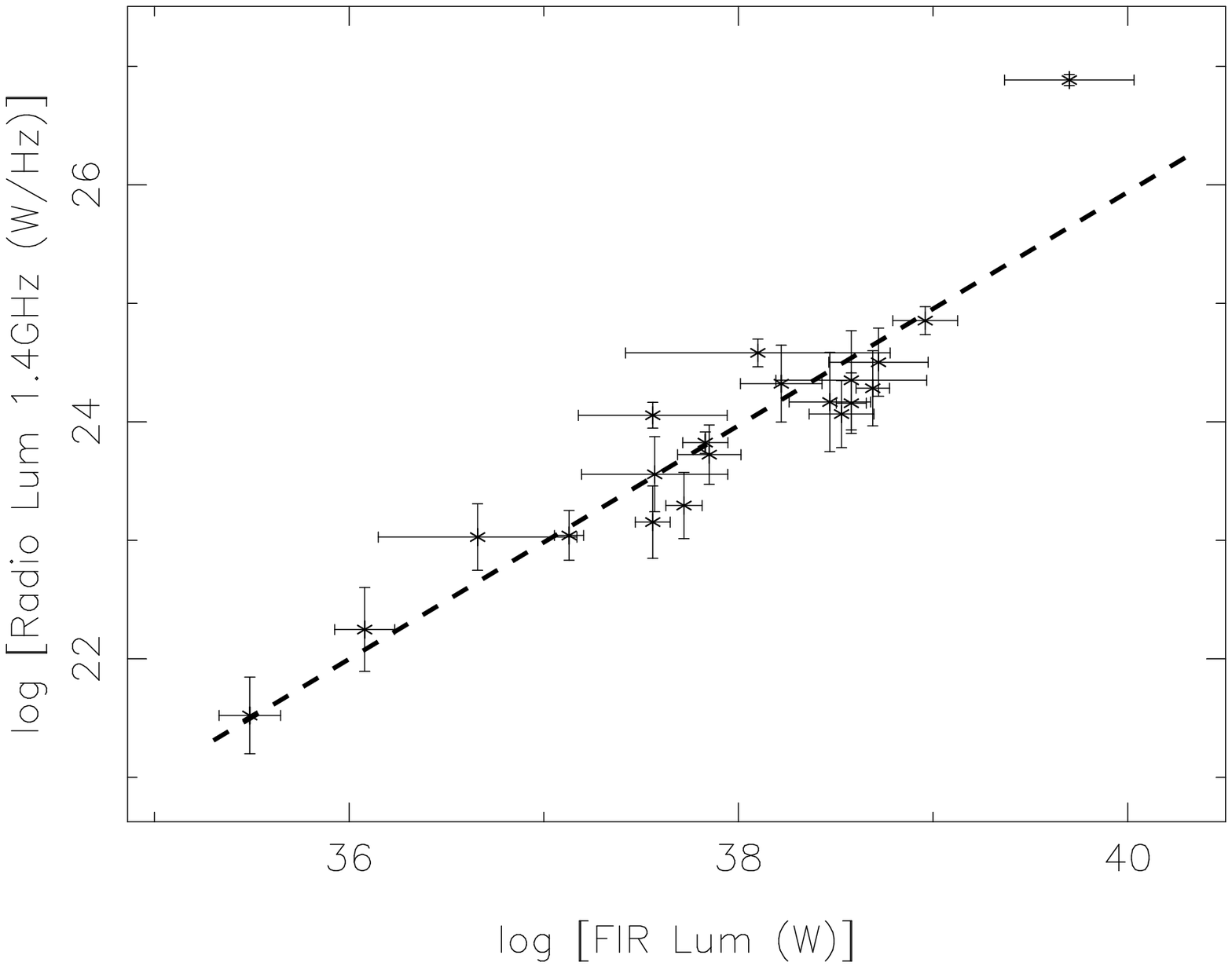}
\caption{Left: MERLIN-VLA contour map (super-imposed on the HST
  gray-scale image) of J123651+621221 -- an
  optically faint dust-obscured starburst system in the HDF with a
  total flux density of $49\mu$Jy (Muxlow et al. 1999). (b) Right: The
  strong correlation between FIR and radio luminosity for high-z
  galaxies detected by both ISO and the WSRT in the region of the HDF-N
  (Garrett 2002)}
\label{fir-rc} 
\end{figure}

Deep VLA observations of the HDF (including the HFF) have been
conducted at both 8.3 and 1.4~GHz (Fomalont et al. 1997, Richards et
al. 1998, Richards 2000). The 8.3~GHz observations reach noise levels
of a few microJy per beam (several times better than the 1.4~GHz
observations) but more sources are actually detected at 1.4~GHz where
the source counts are steeper and the VLA field of view wider. Perhaps
the most ``complete'' radio view of the HDF (see Fig.~\ref{wsrt-im}) is
provided by the WSRT 1.4 GHz observations (Garrett et al.  2000a).
These are sensitive to very extended radio structures, although for the
WSRT (and indeed the VLA), the vast majority of the microJy radio
source population remain barely or completely unresolved at arcsecond
resolution. Combined VLA-MERLIN 1.4 GHz observations with a resolution
of $0.2^{\prime\prime}$ (Muxlow et al.  1999) begin to resolve most of
these sources but the detailed morphology of the microJy radio source
population still remains unknown. The main results of the VLA, MERLIN
and WSRT data can be summarised as follows:

\begin{itemize} 

\item Most of the radio sources are identified with relatively bright
  ($R< 25^{m}$), moderate redshift, optical counterparts -- often
  identified as interacting, irregular or peculiar morphological type 
  (Richards et al. 1998)

\item The 1.4~GHz VLA source sample is steep spectrum in nature (typically
  $\alpha \sim -0.85$ -- Richards 2000). Sources selected by the VLA at 
  8.3 GHz are significantly flatter specutrum ($\alpha \sim -0.35$)
  
\item There is a strong correspondence between the Mid-IR ISO
  detections and the radio detections (see Fig.~\ref{fir-rc} right).
  Indeed the majority of radio sources (after applying appropriate
  k-correction factors) appear to closely follow the FIR-radio
  correlation (Garrett 2002).

\item Of the 91 sources detected with the combined MERLIN-VLA 1.4 GHz
  array (see Fig.~\ref{fir-rc} left), the majority show radio structure on
  {\it sub-galactic} scales (Muxlow et al. 1999). About 50\% of the
  MERLIN-VLA 1.4~GHz detections show extended structure aligned with
  the optical isophotes of the galaxy. 
  
\item The WSRT detects a small but significant population of both star
  forming galaxies and AGN, some of which are resolved by the higher
  resolution VLA and MERLIN observations (Garrett et al. 2000a; Muxlow
  priv. comm).

\item A comparison between the VLA and WSRT 1.4 GHz images (separated
  by several years) shows evidence for significant variability (factors
  of 2 or more) for a few percent of the sub-mJy radio source
  population (presumably low-luminosity AGN).

\item Around 10-20\% of the microJy source population (see
  Fig.~\ref{fir-rc} -left and Richards et al. 1999) are optically faint
  or completely unidentified ($R > 25^{m}$).

\end{itemize} 

\subsection{Summary of Results} 
\label{rad_sum} 

These results, in particular the general correlation between radio and
FIR luminosity, suggest that the radio emission arising in the faint
sub-mJy and microJy radio source population is largely associated with
massive star formation in distant star forming galaxies. However, a
significant fraction of all the sub-mJy sources ($\sim 30$\%) are also
identified with low-luminosity AGN. The remaining 10-20\% of the faint
radio source population are assoiciated with either extremely faint ($R
> 25$) optical identifications, or remain unidentified altogether - even
in the HDF-N itself ($I > 28$). Note that these conclusions are
dominated by the faintest (and therefore more numerous) microJy radio
sources -- for example, the AGN fraction increases rapidly at higher
(sub-mJy) flux density limits. In addition, simply labeling sources as
pure ``starbursts'' or pure ``AGN'' is a little misleading; it is quite
possible (even likely) that both phenomena co-exist in some of these
faint systems. When we label sources in this way, we are only
identifying the {\it dominant} phenomenon that gives rise to the bulk of
the observed radio emission.

Studies of nearby (nuclear) starburst galaxies such as M82 and Arp~220
give us some idea of how the radio emission arises in these star
forming galaxies. According to conventional theories, the chief
ingredients are supernova (SN) events associated with massive star
formation (see Marcaide this lecture series), and in particular, the
global acceleration of cosmic ray electrons via shocks associated with
these events. The total radio luminosity of a ``normal'' galaxy is
therefore a direct measure of the SN event rate, and in turn, the star
formation rate (SFR) of massive stars (e.g. Condon 1992). In this
scenario, the tight correlation between the FIR and radio luminosity of
star forming galaxies is explained by the FIR emission arising from the
absorption and re-radiation (via dust) of the intense uv emission also
associated with massive stars.  By assuming an Initial Mass Function
(e.g. a Salpeter IMF) for the stellar population, and some scaling
factors based on local observations of the Milky Way, radio
observations provide unbiased estimates of the SFR that are largely
unaffected by extinction due to dust. The SFR inferred in this way for
M82 and Arp 220 are $\sim 10$ and $100$~M$_{\odot}$/yr respectively.
The levels of radio and FIR emission observed for the more most distant
radio sources in the HDF implies much higher star formation rates:
$\sim 1000$~M$_{\odot}$/yr.

\subsection{The Optically Faint microJy Radio Source Population} 
\label{scuba_ref} 

There is now good evidence to suggest (e.g. Barger et al. 2000; Chapman
et al. 2001) a close correspondence between the optically faint microJy
radio sources and the faint (SCUBA) sub-mm source population. The bulk
of the observational evidence suggests that these sources are located
at cosmological distances, and are enveloped in thick, opaque dust.
Since it is estimated that a substantial fraction of the total
radiation in the Universe is emitted from these obscured systems, the
detailed nature of these sources is a key topic in astronomy today.
However, the study of these dusty systems is severely hampered by the
fact that they are so difficult to detect in almost all wave-bands,
except the sub-mm and radio domains. It is thought that the sub-mm
emission is associated with FIR emission (again absorption and
re-radiation of uv emission by dust) that is redshifted into the sub-mm
wavelength range.  However, the source of the original uv emission is
unclear -- it could be generated purely by massive stars in dense star
forming regions, purely by embedded AGN, or some mixture of both
phenomena. Similarly in the radio, it is not known whether the
synchrotron emission is generated by massive star formation processes
or by accretion associated with a central AGN.  As we have seen, if
massive star formation is responsible, radio flux densities imply SFRs
$\sim 1000$ M$_{\odot}$/yr, and an inferred global SFR that is
completely ``at odds'' with previous optical and ultraviolet studies
(Haarsma et al.  2001, and references therein).

Distinguishing between the AGN and starburst phenomena in these systems
is clearly of fundamental importance. In the radio, only VLBI provides
sufficient angular resolution to distinguish between the two cases. In
particular, radio emission generated by star formation processes
should be resolved by current (sensitivity limited) VLBI observations.
AGN, on the other hand, should remain considerably more compact and
readily detectable.

\section{Deep Field VLBI observations of the HDF} 
\label{EVN}

On 12-14 November 1999 the EVN conducted the first ``pilot'' VLBI
``blank field'' observations of the radio sky. The field chosen was the
HDF-N -- an area that is just about as ``blank'' and undistinguished as
the radio sky gets.  The brightest source in the $\sim 2$ arcminute
radial field of view was an FR-I radio galaxy with a total WSRT 1.4~GHz
flux density of $\sim 1.6$~mJy. 

\begin{figure}[h]
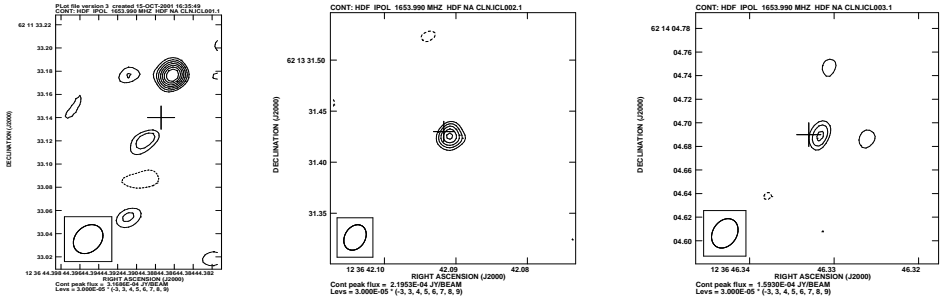

\vspace{4cm}  
\includegraphics{faint_fig5.ps}
\includegraphics{faint_fig6.ps}
\includegraphics{faint_fig7.ps}
\caption{EVN detections in the HDF: the distant
  z=1.01 FRI (left), the z=4.4 dusty obscured starburst hosting a
  hidden AGN (middle) and the
  z=0.96 AGN spiral (right).}
\label{evn_im}
\end{figure}

The data were recorded at a rate of 256 Mbits/sec for 32 hours -- a
sustained capability that is unique to the EVN (and has recently been
extended to 512 Mbits/sec).  Observing in phase-reference mode, a total
of $\sim 14$ hours of ``on-source'' data were collected. With a resolving
beam area 1 million times smaller than the WSRT HDF-N observations (see
section \ref{WSRT}), the EVN imaged an area of about 12
arcmin$^{2}$. Six HDF-N radio sources were thus targeted simultaneously
(using wide-field imaging techniques -- see Garrett et al. 1999). The
final naturally weighted images have an r.m.s. noise level of $\sim
33~\mu$Jy/beam -- much larger than that expected from thermal noise
considerations ($\sim 11~\mu$Jy/beam). The images are probably limited
by the inclusion of poorly calibrated or completely corrupt data 
- difficult to identify in this case, simply by inspection.

Nevertheless, the EVN simultaneously detected three radio sources above
the $165~\mu$Jy ($\sim 5\sigma$) limit, in the inner part of HDF-N
region (see Fig~\ref{evn_im}). These include: VLA~J123644+621133 (a
$z=1.013$, low-luminosity and extremely distant FR-I radio source which
is resolved by the EVN into a core plus hot-spots associated with the
larger scale jet), VLA~J123642+621331 (a dust enshrouded, optically
faint, $z=4.424$ starburst system -- Waddington et al. 1999) and the
faintest detection, VLA~J123646+621404 (a face-on spiral galaxy at
$z=0.96$ with a total EVN flux density of $180~\mu$Jy/beam). The
diversity of optical type is interesting but the real surprise is the
detection of a radio-loud AGN in the dust obscured, $z=4.4$,  starburst
system. This argues that at least some fraction of the optically faint
radio source population harbour hidden AGN (this may be similar to the
same obscured population detected by Chandra). These AGN powered
systems might be quite difficult to detect with SCUBA, assuming the
dust temperatures are higher than that associated with pure star
forming systems. In any case, the detection of this system highlights
the use of VLBI as a powerful diagnostic -- able to distinguish in
principle (via brightness temperature arguments) between radio emission
generated by nuclear starbursts and AGN.

\begin{figure}[h]
\vspace{7cm}  
\includegraphics{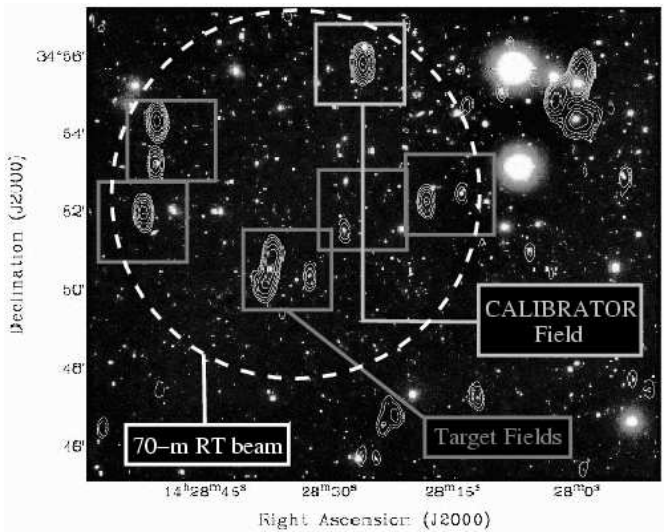}
\caption{A small portion of the NOAO-DWFS with a shallow WSRT 1.4 GHz
  contour map superimposed (image courtesy of Raffaella Morganti, de Vries et
  al. 2001, first radio contour at $30\mu$Jy/beam). The dashed circle
  shows the {\sc FWHM} of a 70-m telescope primary beam. Assuming
  modest r.m.s.  noise levels of $\sim 10~\mu$Jy/beam, 5 ($6\sigma$)
  target fields (of extent $2^{\prime}\times2^{\prime}$) can be
  identified.  These fields are all located within the primary beam,
  and can be correlated simultaneously, and mapped out in their
  entirety. A (faint) VLBI calibrator to the North of the field permits
  accurate and continuous phase calibration to be applied to the target
  fields.}
\label{raff} 
\end{figure}

\section{Future Prospects for deep, wide-field VLBI studies} 

The EVN observations of the HDF suggest that deep, wide-field VLBI
studies are not only possible, but in principle they can deliver
important astronomical results. So far we have only scratched the
surface. In a sense, we are just beginning to appreciate the fact, that
VLBI has reached a sensitivity level where we can expect to detect many
discrete radio sources in a single field of view ({\it irrespective} of
where you point the telescopes!). This is quite a departure from the
traditional role of isolated VLBI observations of very compact, and
often very bright radio sources.

In the short term the use of {\it in-beam} phase-calibration techniques
(Fomalont et al. 1999) should permit us to reach the expected thermal
noise level of only a few microJy/beam (assuming a global VLBI array
operating at 256 Mbits/sec and an on-source integration time of 24
hours). The real advance, however, will be in making full use of the
raw data i.e to map out the primary beam response of individual VLBI
elements in their entirety. Simultaneous multiple-field correlation,
coupled with incredibly fast data output rates, is now being pursued at
the EVN MkIV Data Processor at JIVE.  When complete, this development
will provide astronomers with the ability to image dozens of faint
sub-mJy radio sources -- all observed simultaneously with microJy
sensitivity, full uv-coverage {\it and} milliarcsecond resolution
(Garrett 2000b). Fig.~\ref{raff} summarises the concept of deep,
in-beam, wide-field VLBI but with current sensitivity limits employed.
Large areas of the sky (such as that shown in Fig.~\ref{raff}) are now
being routinely surveyed in great detail by optical and near-IR
instruments. These deep surveys (e.g. the NOAO Deep Wide-Field Survey,
Januzi \& Dey 1999) have the great advantage that they cover enormous
areas of sky (many square degrees) and thus there is always some region
of the survey area that will include an appropriate ``in-beam'' VLBI
calibrator.

What fraction of these target sources can be detected ? The HDF-N
results suggest that about 1/3 of the targets in a randomly selected
field will be AGN, and that most of these will have compact structure.
The remaining distant starburst systems will most likely be resolved by
VLBI (even with microJy sensitivity) -- detecting and imaging these
systems with VLBI scale resolution, must await the construction of a
much more sensitive, next generation radio telescope, such as the SKA
(see Lecture by A.~Kus, this volume). Nevertheless, I suspect that by
the time we return to Castel St. Pietro Therme for the next NATO-ASI
school, the use of VLBI as a deep, wide area survey instrument will
already be well established.

\end{document}